\documentclass[prb,twocolumn,showpacs,amsmath,amssymb,superscriptaddress]{revtex4}

\usepackage[dvips]{graphicx}
\usepackage{graphics}
\usepackage{dcolumn}
\usepackage{bm}
\usepackage[usenames]{color}

\begin{document}

\title{
Local control of magnetocrystalline anisotropy in (Ga,Mn)As
microdevices:  Demonstration in current induced switching}
\author{J. Wunderlich}
\affiliation{Hitachi Cambridge Laboratory, Cambridge CB3 0HE, UK}

\author{A. C. Irvine}
\affiliation{Microelectronics Research Centre, Cavendish Laboratory,
University of Cambridge, CB3 0HE, UK}
\author{J. Zemen}
\affiliation{Institute of Physics ASCR, Cukrovarnick\'a 10, 162 53
Praha 6, Czech Republic}
\author{V. Hol\'{y}}
\affiliation{Charles University, Faculty of Mathematics and Physics,
Department of Electronic Structures, Ke Karlovu 5, 121 16 Prague 2,
Czech Republic}
\author{A. W. Rushforth}
\affiliation{School of Physics and Astronomy, University of
Nottingham, Nottingham NG7 2RD, UK}
\author{E. De Ranieri}
\author{U. Rana}
\affiliation{Microelectronics Research Centre, Cavendish Laboratory,
University of Cambridge, CB3 0HE, UK} \affiliation{Hitachi Cambridge
Laboratory, Cambridge CB3 0HE, UK}
\author{K. V\'yborn\'y}
\affiliation{Institute of Physics ASCR, Cukrovarnick\'a 10, 162 53
Praha 6, Czech Republic}
\author{Jairo Sinova}
\affiliation{Department of Physics, Texas A\&M University, College
Station, TX 77843-4242, USA}
\author{C. T. Foxon}
\author{R. P. Campion}
\affiliation{School of Physics and Astronomy, University of
Nottingham, Nottingham NG7 2RD, UK}
\author{D. A. Williams}
\affiliation{Hitachi Cambridge Laboratory, Cambridge CB3 0HE, UK}
\author{B. L. Gallagher}
\affiliation{School of Physics and Astronomy, University of
Nottingham, Nottingham NG7 2RD, UK}
\author{T. Jungwirth}
\affiliation{Institute of Physics ASCR, Cukrovarnick\'a 10, 162 53
Praha 6, Czech Republic} \affiliation{School of Physics and
Astronomy, University of Nottingham, Nottingham NG7 2RD, UK}
\date{\today}

\begin{abstract}
The large saturation magnetization in conventional dense moment
ferromagnets offers  flexible means of manipulating the ordered
state through demagnetizing shape anisotropy fields but these
dipolar fields, in turn, limit the integrability of magnetic
elements in information storage devices. We show that in a (Ga,Mn)As
dilute moment ferromagnet, with comparatively weaker magnetic dipole
interactions, locally tunable magnetocrystalline anisotropy can take
the role of the internal field which determines the magnetic
configuration. Experiments and theoretical modeling are
 presented for lithographically patterned
 microchannels and the phenomenon is attributed to lattice relaxations across the channels.
The utility of locally controlled magnetic anisotropies is
demonstrated in current induced switching experiments. We report
structure sensitive, current induced in-plane magnetization
switchings well below the Curie temperature at critical current
densities $\sim$10$^{5}$~Acm$^{-2}$. The observed phenomenology
shows signatures of a contribution from domain-wall
spin-transfer-torque effects.
\end{abstract}

\pacs{75.50.Pp, 75.60.Jk, 85.75.-d}

\maketitle
\section{Introduction}
(Ga,Mn)As and related ferromagnetic semiconductors are unique due to
their dilute moment nature and the strong spin-orbit
coupling.\cite{Matsukura:2002_a,Jungwirth:2006_a} Doped with only
$\sim$1-10\% of Mn magnetic moments, the saturation magnetization,
$M_s$, and the magnetic dipole interaction fields are $\sim$100-10
times weaker in these materials than in conventional ferromagnets.
This could make possible dense integration of ferromagnetic
semiconductor microelements with minimal dipolar cross-links.
Despite the low $M_s$ the magnetic anisotropy fields, $H_a$,
routinely reach $\sim$10~mT  due to the large, spin-orbit coupling
induced magnetocrystalline terms.\cite{Dietl:2001_b,Abolfath:2001_a}
The magnetocrystalline anisotropy can, therefore, take the role
normally played by dipolar shape anisotropy fields in the
conventional systems. The combination of appreciable  and tunable
$H_a$ and low  $M_s$ leads to outstanding micromagnetic
characteristics. One particularly important example   is the orders
of magnitude lower critical current in the spin-transfer-torque
magnetization switching\cite{Sinova:2004_b,Chiba:2004_b} than
observed for dense moment conventional ferromagnets, which follows
from the approximate scaling of $j_c\sim H_aM_s$. Critical currents
for domain wall switching of the order 10$^{5}$~Acm$^{-2}$  have
been reported and the effect thoroughly explored in perpendicularly
magnetized (Ga,Mn)As thin film devices at temperatures close to the
Curie
temperature.\cite{Yamanouchi:2004_a,Chiba:2006_a,Yamanouchi:2006_a}

Here we demonstrate that it is possible to locally tune and control
spin-orbit coupling induced magnetocrystalline anisotropies in
(Ga,Mn)As, which is achieved in our devices by lithographically
producing strain relaxation. This is the central result of our work
and it represents the necessary prerequisite for future highly
integrated microdevices fabricated in the dilute-moment
ferromagnets. It also makes possible a range of new studies of
extraordinary magnetotransport and magnetization dynamics effects in
such systems. As a demonstration  we link the achieved local control
of magnetocrystalline anisotropy with a study of current induced
domain wall switching which is currently one of the most hotly
debated areas of theoretical and experimental spintronics
research.\cite{Freitas:1984_a,Yamanouchi:2004_a,Yamaguchi:2004_a,Saitoh:2004_a,Tatara:2004_a,Li:2004_b,Tatara:2004_b,Barnes:2005_a,Thiaville:2005_a,Chiba:2006_a,Yamanouchi:2006_a,Gould:2006_a,Hayashi:2006_a,Thomas:2006_a,Dugaev:2006_a,Xiao:2006_b,Ohe:2006_b,Araujo:2006_b,Duine:2006_a}
We report in-plane domain-wall switchings well below the Curie
temperature at $j_c\sim$10$^{5}$~Acm$^{-2}$ whose characteristics
strongly depend on the locally induced changes of magnetic
anisotropy. The phenomenology of the current induced switching we
observe shows signatures of domain wall spin-transfer-torque
effects.

The paper is organized as follows: In Section~\ref{relax_exp} we
introduce the studied (Ga,Mn)As microstructures and the anisotropic
magnetoresistance (AMR) technique for detecting local magnetization
orientation along the channels.\cite{Hayashi:2006_a} This technique
is particularly useful in dilute moment ferromagnets where direct
imaging methods, such as the magneto-optical Kerr effect, lack the
required sensitivity due to the low $M_s$. Numerical simulations of
the lattice relaxation in the microbars and microscopic calculations
of the corresponding changes of magnetocrystalline anisotropies are
discussed in Section~\ref{relax_theor}. Current induced switching
experiments in our structures with locally controlled anisotropies
are presented in Section~\ref{dw}. A brief summary of the main
results is given in Section~\ref{summary}.

\section{Lattice relaxation and local control of magnetic anisotropy}
\subsection{Experiment}
\label{relax_exp} 
\begin{figure}[h]
\vspace{-1.5cm}
\hspace*{-.8cm}\includegraphics[width=1.2\columnwidth,angle=-0]{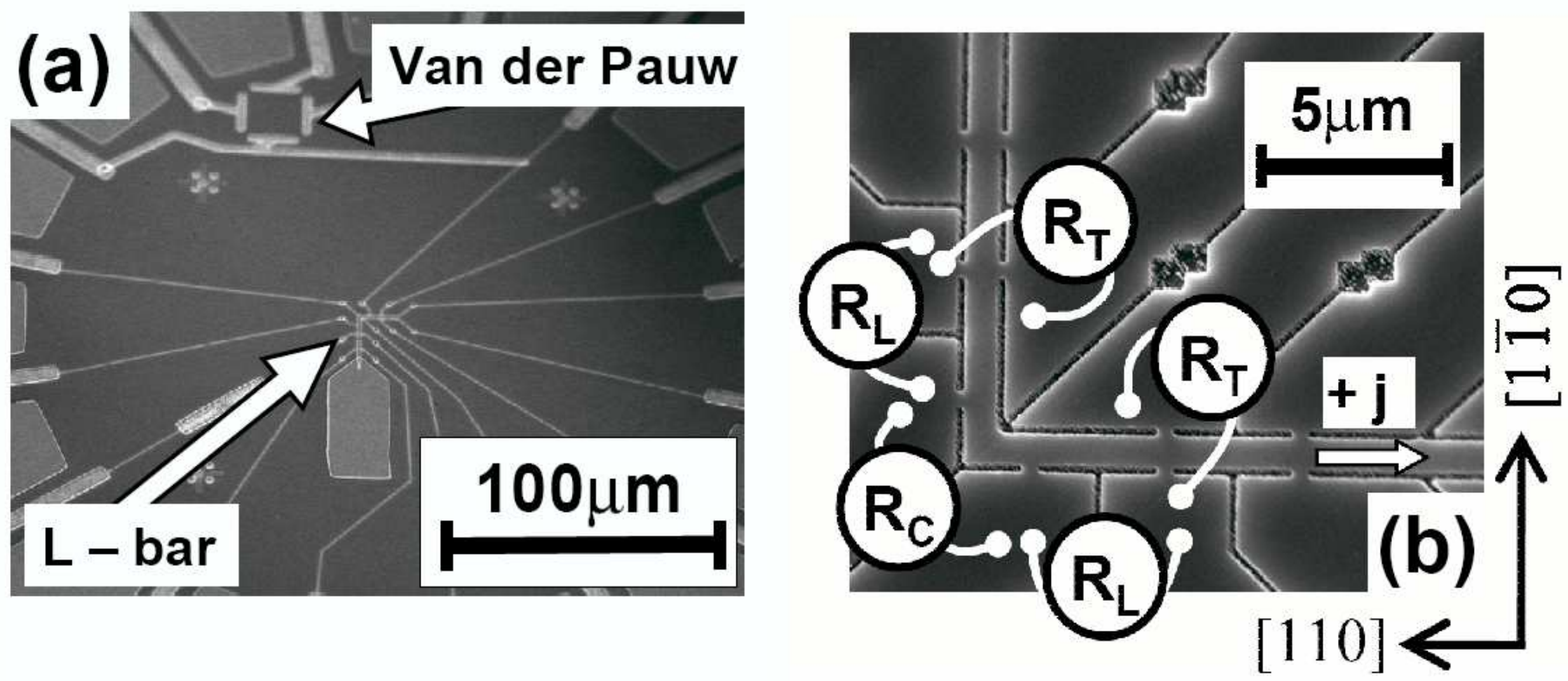}
\vspace*{-1.5cm} 
\caption{(a) Scanning electron micrograph of the
L-shaped microdevice~B and the macroscopic Van der Pauw device. (b)
Detail of the L-shaped microdevice with the longitudinal (L) and
transverse (T) resistance contacts in the bars and the corner (C)
resistance contacts. Positive hole current in the p-type (Ga,Mn)As
is defined to propagate from the [1$\overline{1}$0]-bar to the
[110]-bar.} \label{figure1}
\end{figure}

Fig.~\ref{figure1} shows scanning electron
micrographs of one of the devices studied. The structure consists of
a macroscopic Van der Pauw device and an L-shaped channel patterned
on the same wafer, the arms of which are Hall-bars aligned along the
[1$\overline{1}$0] and [110] directions. The trench-isolation
patterning was done by e-beam lithography and reactive ion etching
in a 25~nm thick Ga$_{0.95}$Mn$_{0.05}$As epilayer, which was grown
along the [001] crystal axis on a GaAs substrate. Results for two
samples are reported: device~A(B) has 4(1)~$\mu$m wide,
80(20)~$\mu$m long Hall bars. Isolated magnetic elements with the
dimensions of these Hall bars and $M_s\sim$50~mT of the
Ga$_{0.95}$Mn$_{0.05}$As would have in-plane shape anisotropy fields
below $\sim$1~mT, which is an order of magnitude lower than the
magnetocrystalline anisotropy fields. In-plane shape anisotropies
are further reduced in our devices as they are defined by narrow
(200nm) trenches with the remaining magnetic epilayer left in place.
The Curie temperature of  100~K was obtained from Arrot plots of
anomalous Hall data. Hole density of 5$\times$10$^{20}$cm$^{-3}$ was
estimated from high-field Hall measurements. At this doping the
compressive strain in the  Ga$_{0.95}$Mn$_{0.05}$As epilayer grown
on the GaAs substrate produces a strong magnetocrystalline
anisotropy which forces the magnetization vector to align parallel
with the plane of the magnetic
epilayer.\cite{Dietl:2001_b,Abolfath:2001_a}

\begin{figure}[h]
\hspace*{-0.8cm}\includegraphics[width=1.2\columnwidth,angle=-0]{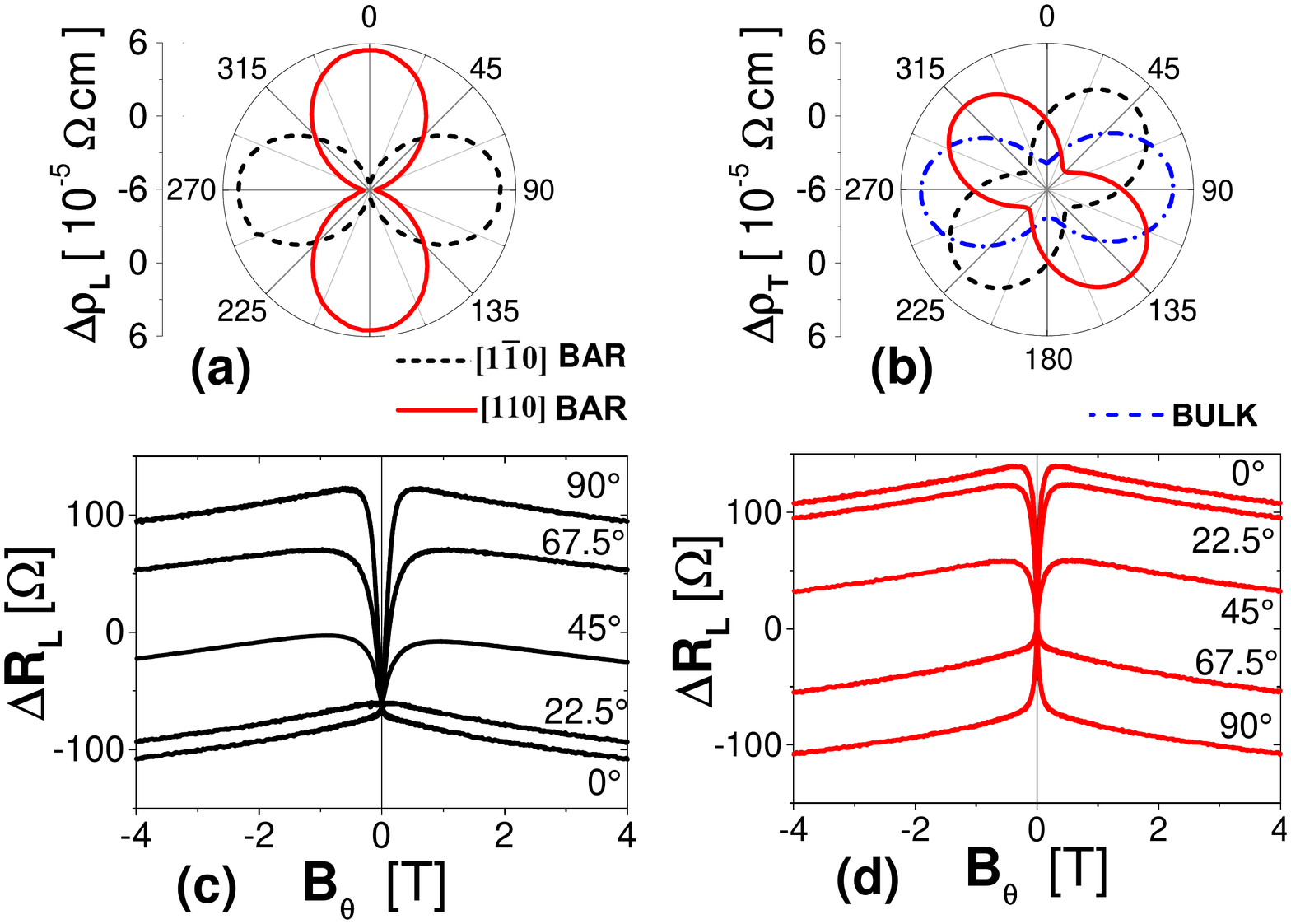}
\caption{Device B longitudinal (a) and transverse (b) AMRs measured
at 4.2~K in a rotating 4~T in-plane field with the field angle
measured from the [1$\overline{1}$0] axis, and bulk transverse AMR
measured in the Van der Pauw device with current lines oriented
along the [010] axis. ($\Delta\rho\equiv \rho-\overline{\rho}$ where
$\overline{\rho}$ is the average value over all angles.) In-plane,
fixed-angle field sweep measurements of the longitudinal
magnetoresistances of the (c) [1$\overline{1}$0]-bar and (d)
[110]-bar bar of device~B. (Same average resistances as in (a)
and (b) are subtracted to obtain $\Delta R$)} \label{figure2}
\end{figure}

Magnetization orientations in the individual microbars are monitored
locally by measuring longitudinal and transverse components of the
 AMR at in-plane magnetic fields. The
magnetization rotation experiments at saturation magnetic field
measured on device~B and on the macroscopic Van der Pauw device are
presented in Figs.~\ref{figure2}(a) and (b). (For the detailed
discussion of the origins of the AMR and microscopic modeling of
this extraordinary magnetoresistance coefficient in (Ga,Mn)As see
Ref.~\onlinecite{Rushforth:2007_a}.)
Examples of magnetoresistance measurements for external magnetic
field sweeps in which the field angle $\theta$, measured from the
[1$\overline{1}$0]~axis, is constant are shown in
Figs.~\ref{figure2}(c) and (d). The strongly $\theta$-dependent
low-filed magnetoresistance is attributed to magnetization
rotations. At high fields, the magnetoresistance becomes purely
isotropic, i.e., the differences between resistances for different
angles $\theta$ become independent of the magnitude of the external
field. This property and the much smaller magnitude of the isotropic
magnetoresistance compared to the low-field anisotropic
magnetoresistance allows us to use the high-field measurements in
Figs.~\ref{figure2}(a),(b) for determining the one to one
correspondence between a change in the low-field resistance and a
change in magnetization orientation. Note that the 45$^{\circ}$
phase shift between the longitudinal and transverse AMR traces (see
Figs.~\ref{figure2}(a),(b)) allows us to determine unambiguously the
change in the magnetization angle if both resistance components are
measured simultaneously. The technique of detecting magnetization
rotations via AMR measurements is exploited in Section~\ref{dw}
where we compare field induced and current induced magnetization
switchings. Importantly, the multiterminal design of our L-shaped
microbars also allows to apply this electrical measurement of
magnetization angle locally at the corner and at different parts of
the L-shaped Hall bars and, therefore, to track the propagation of
domain walls if present in the system.

In this section we use the fixed-$\theta$ magnetoresistance
measurements to first determine local magnetic anisotropies in the
individual microbars. Values of $\theta$ corresponding to easy-axis
directions have the smallest low-field magnetoresistance. For values
of $\theta$  not corresponding to easy-axis directions the
magnetization undergoes a (partially) continuous rotation at low
fields resulting in different orientations, and hence different
measured resistances, at saturation and remanence. We find that the
technique can be used to determine the easy-axis directions within
$\pm 1^{\circ}$.

\begin{figure}[h]
\vspace{-1cm}
\hspace*{-4cm}\includegraphics[width=1.8\columnwidth,angle=0]{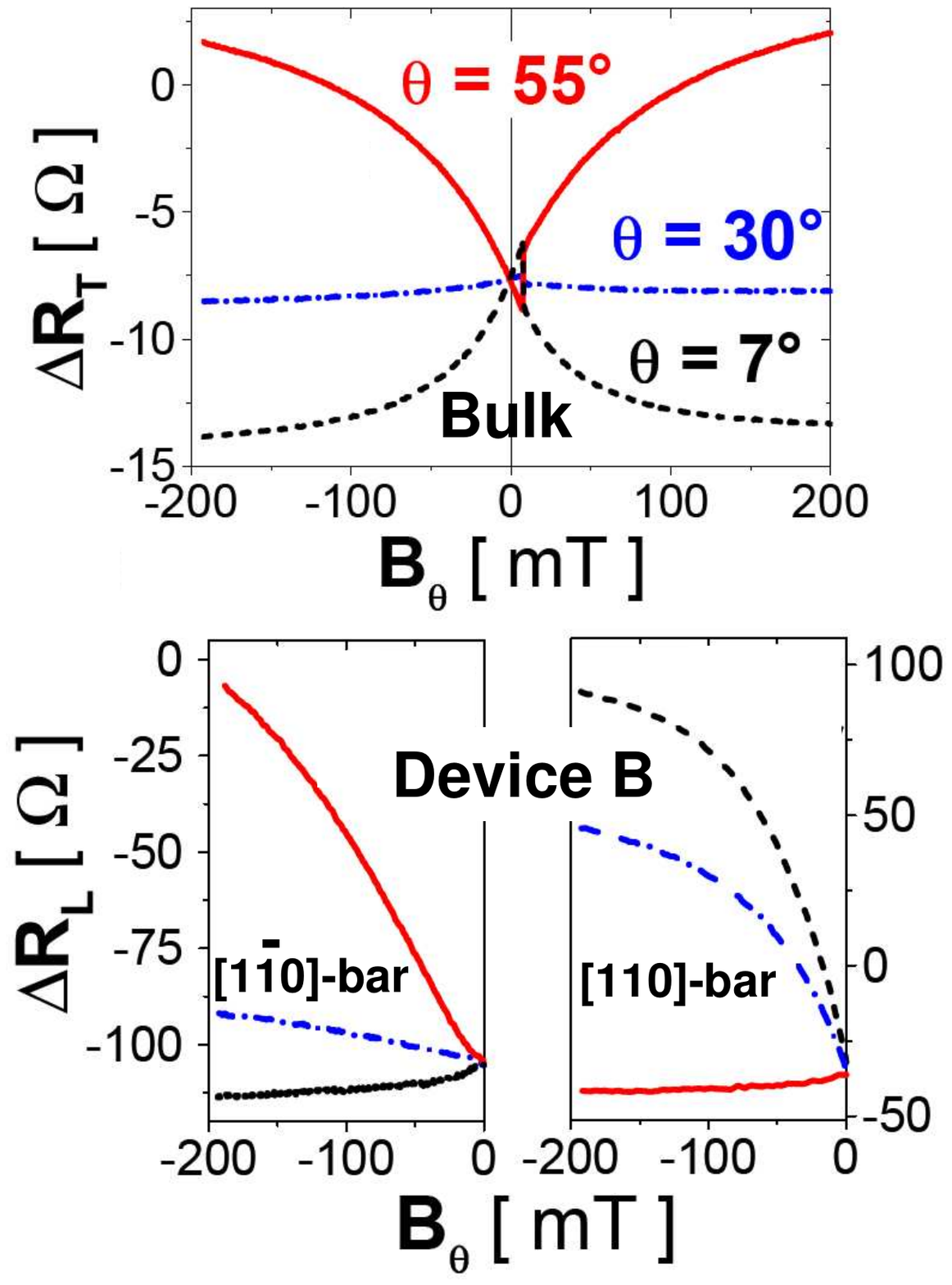}
\vspace{-1cm} 
\caption{Comparison of the low-filed measurements at
4.2~K  of the transverse resistance in the bulk Van de Pauw device
(upper panel) and of the longitudinal resistance of the
[1$\overline{1}$0] and [110]-bar in device B (lower panels).}
\label{figure3}
\end{figure}

The effect of microfabrication on the magnetic anisotropy is
apparent in Fig.~\ref{figure3}. In the bulk, magnetization angle
$30^{\circ}$ corresponds to an easy-axis while $7^{\circ}$ and
$55^{\circ}$ are significantly harder. For device~B, $7^{\circ}$ is
 an easy-axes in the [1$\overline{1}$0]-bar and $55^{\circ}$ is an easy-axis in the [110]-bar.
All easy-axes found in devices~A and B and in the bulk are
summarized in Tab.\ref{tab}. The bulk material has the cubic
anisotropy of the underlying zincblende structure plus an additional
uniaxial [1$\overline{1}$0] anisotropy as is typical (Ga,Mn)As
epilayers.\cite{Sawicki:2004_a} This results in two easy-axes tilted
by 15$^{\circ}$ from the [100] and [010] cube edges towards the
[1$\overline{1}$0] direction. In the microdevices, the easy-axes are
rotated from their bulk positions towards the direction of the
respective bar and the effect increases with decreasing bar width.

\begin{table}[h]
\begin{tabular}{c|c|c|c|c|c}
sample & bulk & A [1$\overline{1}$0] & A [110] & B [1$\overline{1}$0] & B [110]   \\
\hline easy-axis angle & $\pm30^{\circ}$ & $\pm15^{\circ}$ &
$\pm36^{\circ}$ & $+7^{\circ}$,$-8^{\circ}$ &
$+55^{\circ}$,$-63^{\circ}$  \\
\end{tabular}
\caption{Easy-axes angles, measured from the [1$\overline{1}$0]
crystal direction, determined by magnetoresistance measurements in
the macroscopic Van der Pauw device (bulk) and in the
[1$\overline{1}$0] and [110]-bars of the L-shaped devices~A and B.}
\label{tab}
\end{table}

\subsection{Theory}
\label{relax_theor} The local changes in the magnetocrystalline
anisotropy can be understood in the following way.
Ga$_{0.95}$Mn$_{0.05}$As epilayers grown on GaAs substrate are
compressively strained in the (001) plane with the typical value of
the strain parameter
$f\equiv(a^{\ast}_{GaMnAs}-a^{\ast}_{GaAs})/a^{\ast}_{GaAs}\approx
0.2-0.3$\%,  where $a^{\ast}_{GaAs}$ and $a^{\ast}_{GaMnAs}$ are the
lattice parameters of the cubic fully relaxed GaAs and (Ga,Mn)As
film, respectively. With the (Ga,Mn)As material removed in the
trenches along the bars, the lattice can relax in the transverse
direction and the corresponding extension can be roughly estimated
as $f t/w\sim 0.01\%$, where $t=25$~nm is the thickness of the
(Ga,Mn)As film and $w$ is the bar width.

\begin{figure}[h]
\hspace*{-1.2cm}\includegraphics[width=1.3\columnwidth,angle=0]{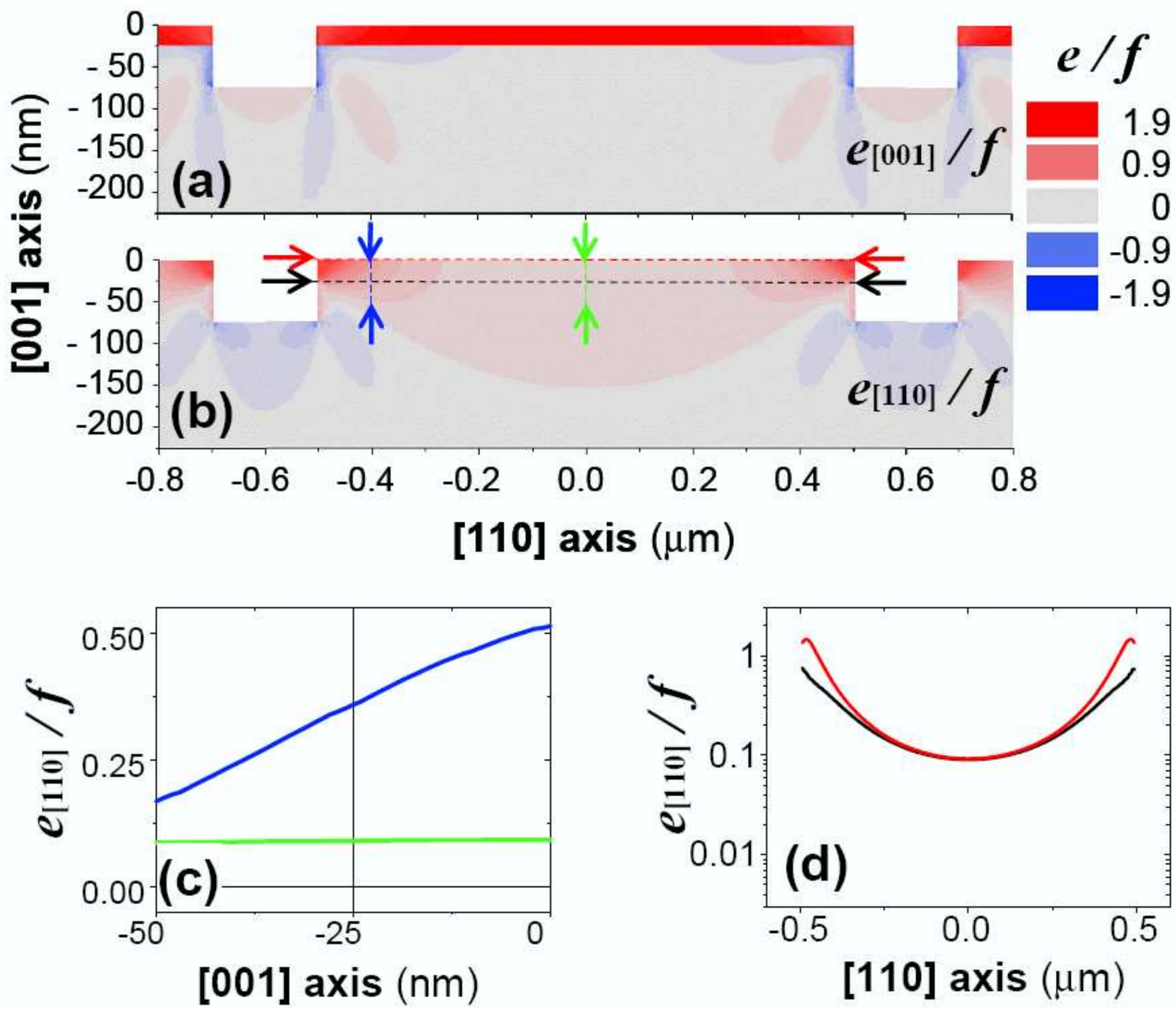}
\caption{Numerical simulations of lattice parameters in the 1~$\mu$m
wide [1$\overline{1}$0]-bar of device B defined by 200~nm wide and
75~nm deep trenches in the 25~nm thick (Ga,Mn)As film on a GaAs
substrate. (a) Strain component along the [001]-axis  with respect
to the lattice parameter of a fully relaxed cubic GaAs,
$e_{[001]}=(a_{[001]}-a^{\ast}_{GaAs})/a^{\ast}_{GaAs}$. The
epitaxial growth induced strain  parameter $f$ is defined as,
$f=(a^{\ast}_{GaMnAs}-a^{\ast}_{GaAs})/a^{\ast}_{GaAs}$ where
$a^{\ast}_{GaMnAs}>a^{\ast}_{GaAs}$ is the lattice parameter of the
cubic fully relaxed (Ga,Mn)As film. (b) Same as (a) for in-plane
strain component $e_{[110]}$ in the direction perpendicular to the
bar orientation. (c) and (d) Strain components $e_{[110]}$ along
different cuts through the [001]-[110] plane. The cuts and the
corresponding $e_{[110]}/f$ curves are highlighted by colored arrows
in (b) and the corresponding color coding of curves in (c) and (d).}
\label{figure4}
\end{figure}

On a quantitative level, the strength of the lattice relaxation in
the microbars is obtained from numerical elastic theory simulations
for the realistic sample geometry. (GaAs values of the elastic
constants are considered for the whole wafer including the
Ga$_{0.95}$Mn$_{0.05}$As epilayer.) Results of such calculations are
illustrated  in Fig.~\ref{figure4} for the [1$\overline{1}$0]-bar of
device B. In panel (a) we show the strain component along the
growth-direction [001]-axis with respect to the lattice parameter of
a fully relaxed cubic GaAs,
$e_{[001]}=(a_{[001]}-a^{\ast}_{GaAs})/a^{\ast}_{GaAs}$. Since all
strain components scale linearly with $f$ we plot $e_{[001]}/f$. The
figure highlights the growth induced lattice matching strain;
because of the in-plane compression of the (Ga,Mn)As lattice the
elastic medium reacts by expanding the lattice parameter in the
growth direction, as  compared to $a^{\ast}_{GaMnAs}$, i.e.,
$e_{[001]}/f>1$.

Within the plane, the lattice can relax only in the direction
perpendicular to the microbar orientation. The corresponding strain
component, calculated again with respect to the GaAs, is plotted in
Fig.~\ref{figure4}(b) over the entire cross-section of device B and,
in Figs.~\ref{figure4}(c) and (d), along various cuts through the
[001]-[110] plane. While in the center of the bar the in-plane
relaxation is relatively weak, i.e. the lattice parameter remains
similar to that of the GaAs substrate, the lattice is strongly
relaxed near the edges of the bar. Averaged over the entire
cross-section of the (Ga,Mn)As bar we obtain relative in-plane
lattice relaxation of several hundredths of a per cent, i.e., of the
same order as estimated by the $f t/w$ expression. The microscopic
magnetocrystalline energy calculations discussed in the following
paragraphs confirm that these seemingly small lattice distortions
can fully account for the observed easy-axis rotations in the
strongly spin-orbit coupled (Ga,Mn)As.

Our microscopic calculations of the magnetization angle dependent
total energies are based on combining the six-band ${\bf k}\cdot
{\bf p}$ description of the GaAs host valence band with
kinetic-exchange model of the coupling to the local Mn$_{\rm Ga}$
$d^5$-moments.\cite{Dietl:2001_b,Abolfath:2001_a} The theory is well
suited for the description of spin-orbit coupling phenomena in the
top of the valence band whose spectral composition and related
symmetries are dominated, as in the familiar GaAs host, by the
$p$-orbitals of the As sublattice. The ${\bf k}\cdot {\bf p}$
modeling also provides straightforward means of accounting for the
effects of lattice strains on the (Ga,Mn)As band
structure.\cite{Dietl:2001_b,Abolfath:2001_a} (As in the above
macroscopic  simulations we assume that the elastic constants in
(Ga,Mn)As have the same values as in GaAs.) This theory, which uses
no adjustable free parameters, describes accurately the sign and
magnitude of the AMR data in
Fig.~\ref{figure2}.\cite{Rushforth:2007_a} It has also explained the
previously observed transitions between in-plane and out-of-plane
easy magnetization orientations in similar (Ga,Mn)As epilayers grown
under compressive and tensile strains and provided a consistent
account of the signs and magnitudes of corresponding AMR
effects.\cite{Jungwirth:2006_a}

For the  modeling of the magnetocrystalline energy of the microbars
we assume homogeneous strain in the (Ga,Mn)As layer corresponding to
the average value of $e_{[110]}$ obtained in the macroscopic elastic
theory simulations.  The input parameters of the microscopic
calculations\cite{Dietl:2001_b,Abolfath:2001_a} are then strain
components, related to the fully relaxed cubic (Ga,Mn)As lattice, in
the [100]-[010]-[001] ($x-y-z$) coordinate system which are given
by:
\begin{eqnarray}
e_{ij} &=& \left(\begin{array}{ccc} e_{xx} & e_{xy}& 0 \\
e_{yx} & e_{yy}& 0 \\
0 & 0 & e_{zz} \\
\end{array}\right)\nonumber \\ \nonumber \\
&=& \left(\begin{array}{ccc} \frac{e_{[110]}}{2} - f& \pm
\frac{e_{[110]}}{2}& 0 \\
\pm \frac{e_{[110]}}{2} &
\frac{e_{[110]}}{2} - f & 0 \\
0 & 0 & e_{[001]} - f  \\
\end{array}\right)\;,
\end{eqnarray}
where $\pm$ corresponds to the [1$\overline{1}$0]-bar and [110]-bar
respectively.

\begin{figure}[h]
\hspace*{-0cm}\includegraphics[width=1\columnwidth,angle=-0]{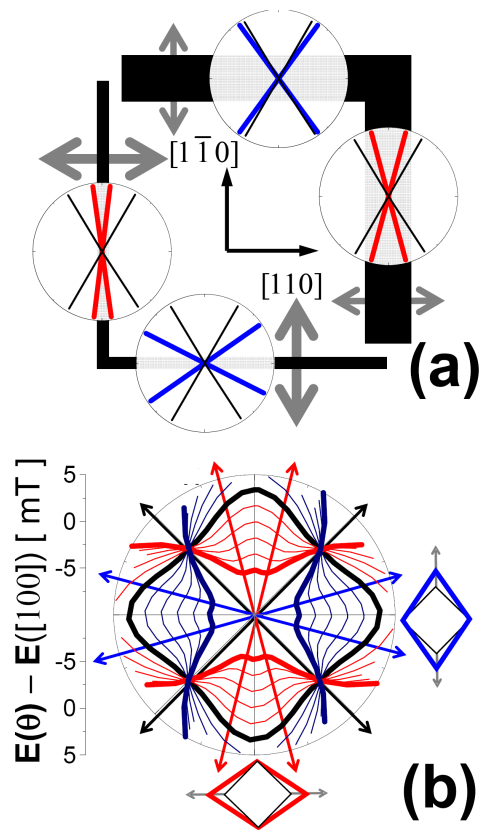}
\caption{(a) Schematics of the easy-axes orientations
in the [1$\overline{1}$0] and [110]-bars of the L-shaped devices A
and B. Arrows indicate the direction and strength of the patterning
induced lattice relaxation. (b) Theoretical magnetocrystalline
energies as a function of the in-plane magnetization angle for zero
shear strain (black line), for $e_{xy}=0.004,..,0.02$\% (red lines)
corresponding to lattice extension along [110] axis, and for
$e_{xy}=-0.004,..,-0.02$\% (blue lines) corresponding to lattice
extension along  [1$\overline{1}$0] axis. The magnetic easy-axes at
$e_{xy}=0$, 0.02\% and -0.02\% are highlighted by  black, red, and
blue arrows, resp. Lattice deformations breaking the
[1$\overline{1}$0]-[110] symmetry of the microscopic
magnetocrystalline energy profile are illustrated by the
diamond-like unit cells extended along [110] axis for the
[1$\overline{1}$0]-bar (red diamond) and along the
[1$\overline{1}$0] axis for the [110]-bar (blue diamond). }
\label{figure5}
\end{figure}

In Fig.~\ref{figure5}(b) we plot calculated magnetocrystalline
energies as a function of the in-plane magnetization angle  for
$f=0.3$\% and $e_{xy}$ ranging from zero (no in-plane lattice
relaxation) to typical values expected for the
[1$\overline{1}$0]-bar ($e_{xy}>0$) and for the [110]-bar
($e_{xy}<0$). Consistent with the experiment, the minima at [100]
and [010] for $e_{xy}=0$ move towards the
 [1$\overline{1}$0] direction for lattice expansion along [110] direction ($e_{xy}>0$) and
 towards the [110] direction for lattice expansion along [1$\overline{1}$0] direction ($e_{xy}<0$).
 Note that the asymmetry between
 experimental easy-axes rotations in the two bars is due to the a [110]-uniaxial component present already in
 the bulk material whose microscopic
 origin is not known but can be modeled\cite{Sawicki:2004_a} by an intrinsic (not induced by
micropatterning) strain
 $e^{bulk}_{xy}\sim+0.01$\%.

\section{Demonstration in current induced switching}
\label{dw}
 The L-shaped geometry of our devices is well suited for a
 systematic study of the link between the locally adjusted magnetic
 anisotropies in the individual microbars and their current induced
 switching characteristics. Apart from the distinct
 magnetocrystalline anisotropy fields, the two bars in each device
 have identical material parameters and lithographical dimensions.
 They can also be expected to share a common domain-wall nucleation
 center at the corner of the L-shaped channel since in this region
 the lattice relaxation effects and the corresponding enhancement of
 the magnetocrystalline anisotropies are less pronounced. Apart from this
effect, the domain wall nucleation at the corner can be expected to be supported by
an enhanced current induced heating in this part of the device.

\begin{figure}[h]
\hspace*{-1.3cm}\includegraphics[width=1.3\columnwidth,angle=0]{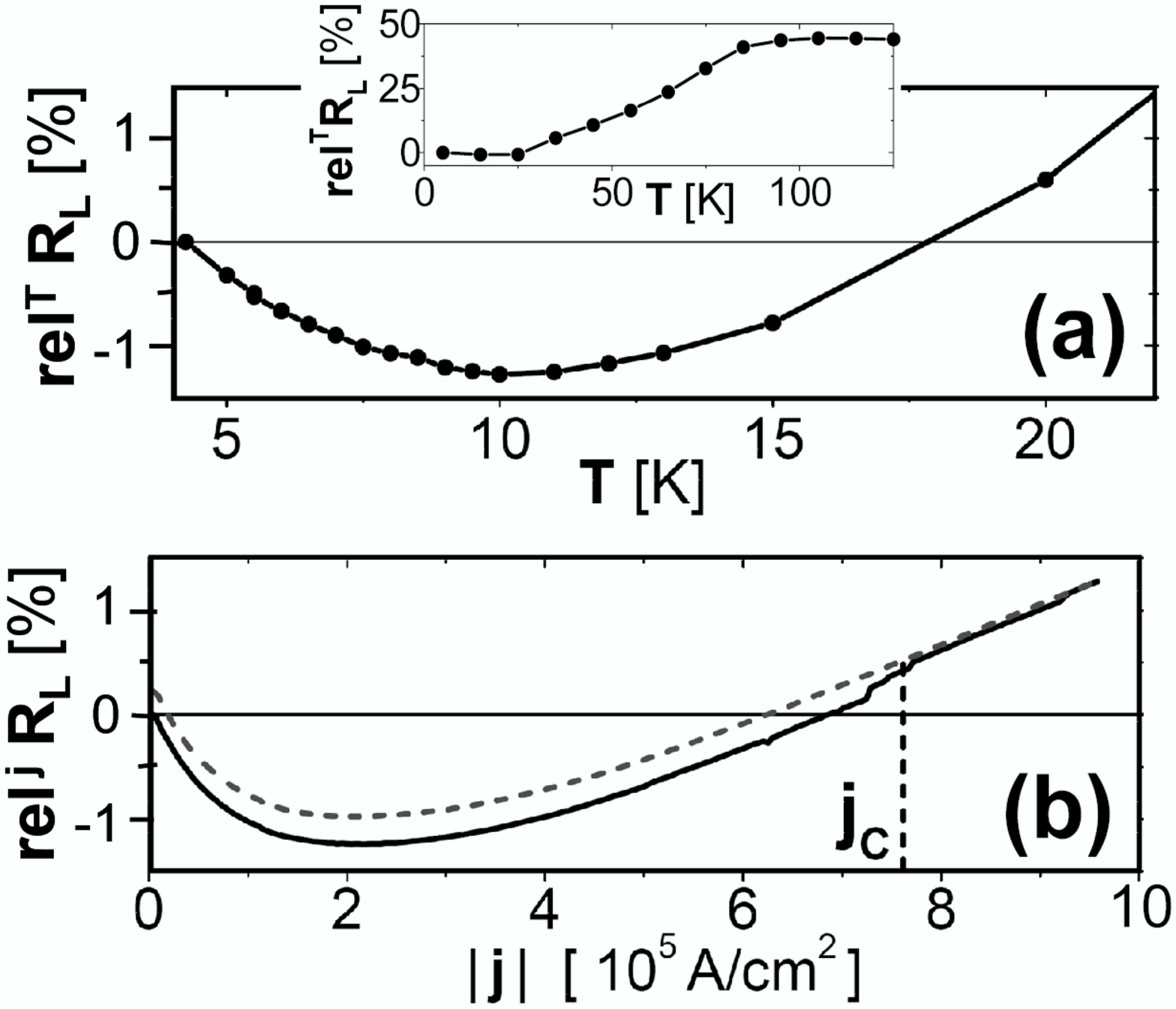}
\caption{ (a) Temperature dependence of
 ${\rm rel}^T R_L \equiv[R_L(T)-R_L(4.2)]/R_L(4.2)$
 at current
 density 10$^{3}$~Acm$^{-2}$. For completeness, ${\rm rel}^T R_L$ over a wide range of temperatures below and above
the Curie temperature is shown in the inset. (b) First (solid line) and second (dashed line) current ramps
 at -4~mT field applied along  $\theta=7^{\circ}$; relative resistances are plotted with respect to the
 zero-current resistance in the first ramp. Switching at $j_c\approx-7.5\times$10$^{5}$~Acm$^{-2}$ is marked.
 }
\label{figure6}
\end{figure}

\begin{figure}[h]
\vspace*{-.5cm}
\hspace*{-.2cm}\includegraphics[width=1.05\columnwidth,angle=-0]{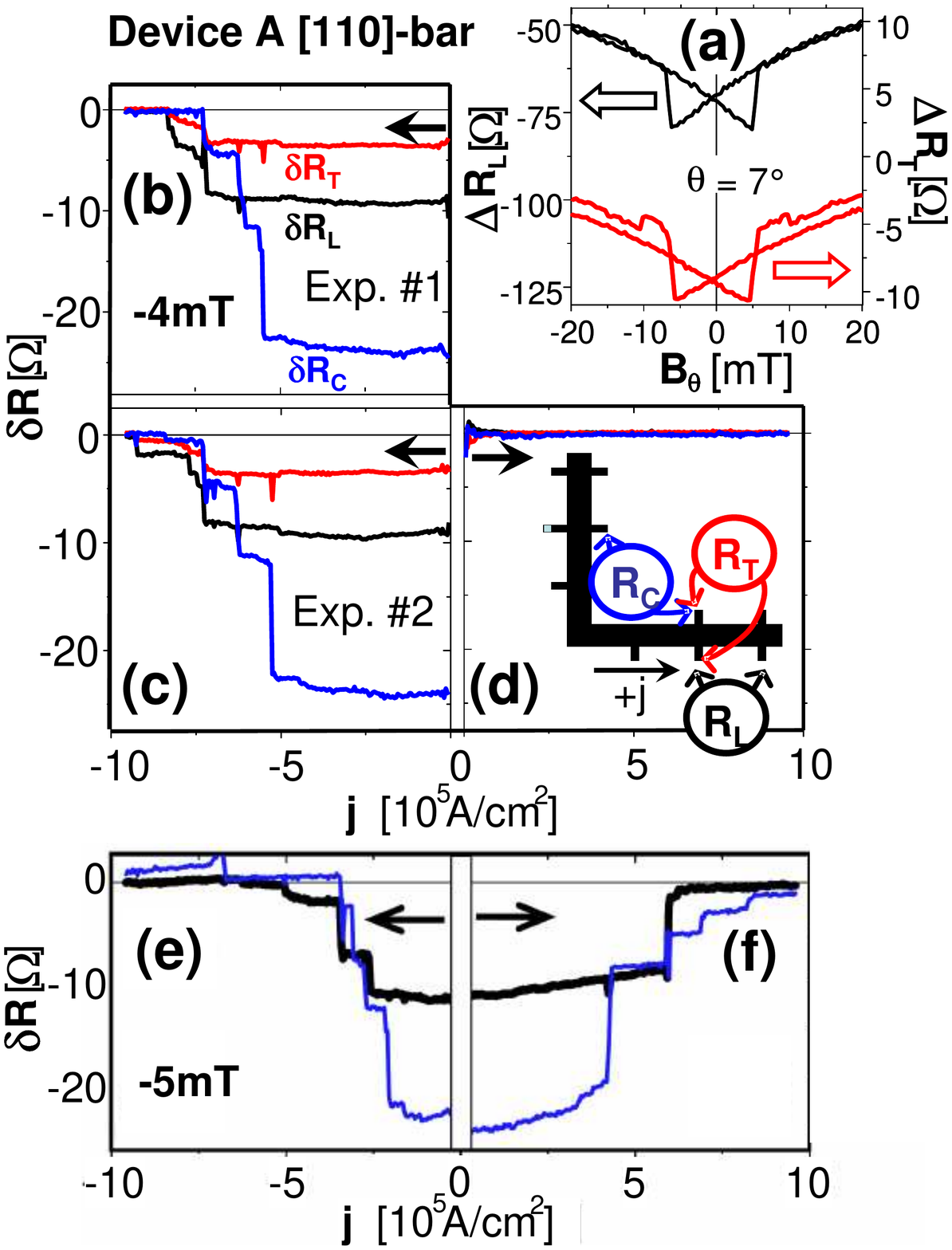}
\vspace{-.5cm} 
\caption{
 (a) Field-sweep measurements at $\theta=7^{\circ}$  in the [110]-bar  of device~A.
 (b) Differences between the first and second negative current ramps for the
 longitudinal (black lines) and transverse (red lines) resistance in the [110]-bar
 and in the corner (blue lines) of device~A at -4~mT external field applied along
$\theta=7^{\circ}$. Arrows indicate the current ramp direction.
 (c) Same as (b) for the second independent experiment. (d) Same as (b) and (c) for
 positive current ramps. The inset shows contacts used for measurements of $R_L$, $R_T$, and $R_C$ in all panels.
 (e),(f) -5~mT field assisted current induced switching experiments.}
\label{figure7}
\end{figure}

 The basic
 phenomenology of current induced switchings that we observe in all
 our L-shaped microbars is illustrated in Figs.~\ref{figure6} and
 \ref{figure7}. The particular field-assisted switching data plotted
 in the figures were measured in the [110]-bar of device A at
 $\theta=7^{\circ}$. At this off-easy-axis angle the current induced
 switching can be easily induced and  detected due to  the
 hysteretic bistable character of the low field magnetization and
 the clear AMR signal upon reversal (see Fig.~\ref{figure7}(a)). We
 start with assessing the role of heating in the current induced
 switching experiments. Figs.~\ref{figure6}(a) and (b) compare the
 temperature dependence of the longitudinal resistance at low
 current density (10$^{3}$~Acm$^{-2}$) with the dependence on
 current density measured in liquid helium. As seen from the plots,
 the maximum current density of 1$\times$10$^{6}$~Acm$^{-2}$ used in
 the experiments corresponds to heating the sample by  approximately
 20~K, which is well below the Curie temperature of 100~K.
 Nevertheless, a suppression due to heating of the effective barrier
 between metastable and stable states and thermally induced
 reversals are possible near the switching fields and these effects have to be considered
 when analyzing the current induced switching experiments below.

 The measurements presented in Figs.~\ref{figure7}(b)-(f) were
 performed by first applying a saturation field  and then reversing
 the field and setting it  to a value close to but below the
 switching field in the field-sweep experiment (see Fig.~\ref{figure7}(a)). Then, the first
 current ramp was applied which triggered the reversal, followed by
 subsequent control current ramps of the same polarity which showed
 no further changes in the magnetization. Constant current sweep
 rate of  $5\times 10^{4}$~Acm$^{-2}$s$^{-1}$ was used in all
 experiments. In Figs.~\ref{figure7}(b)-(f) we plot the difference, $\delta
 R$, between  resistances of the first and the subsequent current
 ramps. We note that no switchings were observed in these experiments
up to the highest applied currents in the [1$\overline{1}$0]-bar. In this bar with the stronger
magnetocrystalline anistropy, the magnitude of the low current (10$^{3}$~Acm$^{-2}$)
switching field at $\theta=7^{\circ}$ is
$\approx 8$~mT, as compared to the $\approx 5.5$~mT switching field in the [110]-bar.

 First we discuss data in Fig.~\ref{figure7}(b) and (c) taken at  -4~mT external field and negative current ramps.
 The two independent experiments (panels (b) and (c) respectively)
 performed at nominally identical conditions demonstrate
 the high degree of reproducibility achieved in our devices. This
 includes the step-like features which we associate with domain wall
 depinning/pinning events preceding full reversal. To understand this process in more detail we
 complement the longitudinal (black curve) and transverse (red curve) resistance measurements in the [110]-bar
 with the resistance measurements at the corner (blue curve) of the L-shaped channel. The schematic plot
 of the respective voltage probes is shown in the inset. The
 first magnetization switching event at $j\approx-5\times$10$^{5}$~Acm$^{-2}$ is detected by the step in the $\delta R_C$
 signal, i.e., occurs
 in the corner region between the $R_C$ contacts. For current
 densities in the range between
 $j\approx-5\times$10$^{5}$~Acm$^{-2}$ and
 $j\approx-6\times$10$^{5}$~Acm$^{-2}$ the domain wall remains
 pinned in the corner region. The next domain wall propagation and pinning event in $\delta R_C$ is
 observed between $j\approx-6\times$10$^{5}$~Acm$^{-2}$
and  $j\approx-7\times$10$^{5}$~Acm$^{-2}$ and for $|j|>7\times$10$^{5}$~Acm$^{-2}$ the region
 between the $R_C$ contacts is completely reversed. The depinning events
 at $j\approx-5\times$10$^{5}$~Acm$^{-2}$ and
 $j\approx-6\times$10$^{5}$~Acm$^{-2}$ are also registered by the
 $R_L$ and $R_T$ contacts through noise spikes in the respective $\delta R_L$ and $\delta R_T$
signals. However, beyond these spikes, $\delta R_L$ and $\delta R_T$
remain constant for $|j|<7\times$10$^{5}$~Acm$^{-2}$ indicating that
the domain wall has not reached the section of the [110]-bar between
the $R_L$ contacts at these current densities. Constant $\delta R_C$
and step-like changes in $\delta R_L$ and $\delta R_T$
at$|j|>7\times$10$^{5}$~Acm$^{-2}$ are signatures of the domain wall
leaving the corner section and entering the part of the [110]-bar
between the $R_L$ contacts. The reversal of this part is completed
at $j\approx-8\times$10$^{5}$~Acm$^{-2}$. Note that both the
 $\delta R_L$, averaging over the whole bar between the longitudinal
 contacts, and the $\delta R_T$, reflecting the local structure near
 the respective transverse contacts, show switching at the same
 current and the sense and magnitude of the overall change in $\delta R_L$ and $\delta R_T$
 are
 consistent with  those observed in the field sweep measurement (see
 Fig.~\ref{figure7}(a)). This indicates that the contacts have a
 negligible effect on the anisotropy in this bar and allows us to
 unambiguously determine the magnetization angles of the initial
 state, $39\pm1^{\circ}$, and of the final state, $211\pm1^{\circ}$. This
 -4~mT field assisted current induced switching is not observed
 at positive current ramps up to the highest experimental current density of
$j=1\times$10$^{6}$~Acm$^{-2}$
 which indicates that
 spin-transfer-torque effects can be contributing to the reversal. Note also that
 the domain wall propagates in the
 direction opposite to the applied hole current, in agreement  with
 previous spin-transfer-torque studies of
 perpendicularly magnetized (Ga,Mn)As films.\cite{Chiba:2006_a} (The
 anomalous direction of the domain wall propagation is assigned to the
 antiferromagnetic alignment of hole spins with respect to the total
 moment in
 (Ga,Mn)As.\cite{Yamanouchi:2004_a,Chiba:2006_a,Yamanouchi:2006_a})

A suppression of the role of the spin-transfer-torque relative to
the thermally assisted switching mechanism is expected at fields
closer to the coercive field. The data taken at -5~mT field shown in
 Fig.~\ref{figure7}(e) and (f) are fully consistent with this expectation.
 Current induced switchings are observed here at lower critical currents and for both
 current polarities. Nevertheless, the asymmetry between the negative and
 positive critical currents is still apparent and consistent with a
 picture of cooperative effects of heating and spin-transfer-torque
 for negative currents and competing effects of the two mechanisms
 for positive currents.

\begin{figure}[h]
\vspace*{-1cm}
\hspace*{-0.5cm}\includegraphics[width=1.1\columnwidth,angle=-0]{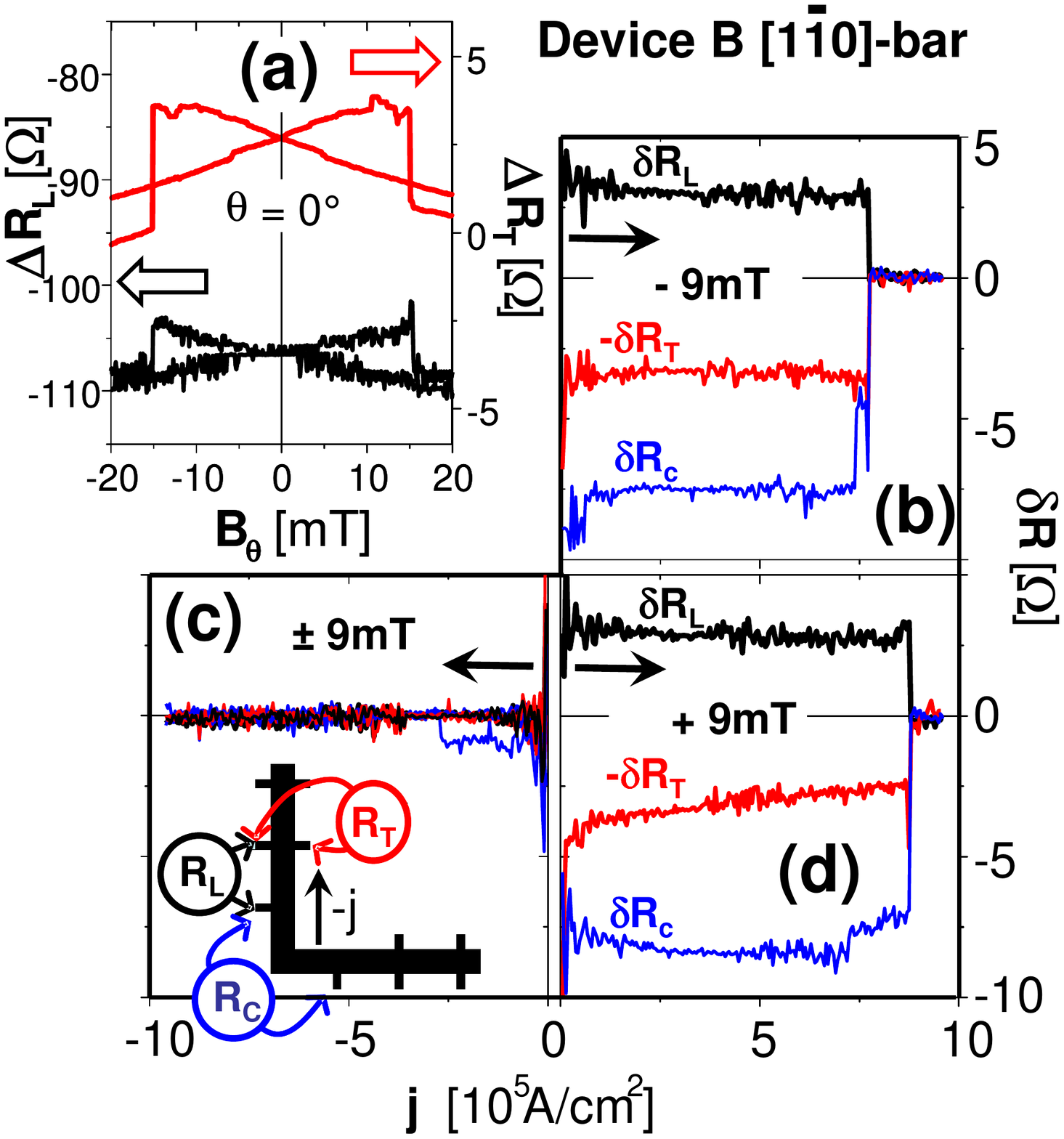}
\vspace{-3cm}
\caption{(a) Field-sweep measurements at
$\theta=0^{\circ}$  in the [1$\overline{1}$0]-bar of device~B. (b)
Difference between the first and second positive current ramps in
the [1$\overline{1}$0]-bar of device~B at -9~mT field applied along
$\theta=0^{\circ}$. Note that $-\delta R_T$ is plotted for clarity. (c) Same as (b) at negative current ramps at
$\pm 9$~mT. The inset shows contacts used for measurements of $R_L$
and $R_T$ in all panels.(d) Same as (b) at $+9$~mT field.}
\label{figure8}
\end{figure}

\begin{figure}[h]
\vspace*{-1cm}
\hspace*{-.4cm}\includegraphics[width=1.1\columnwidth,angle=-0]{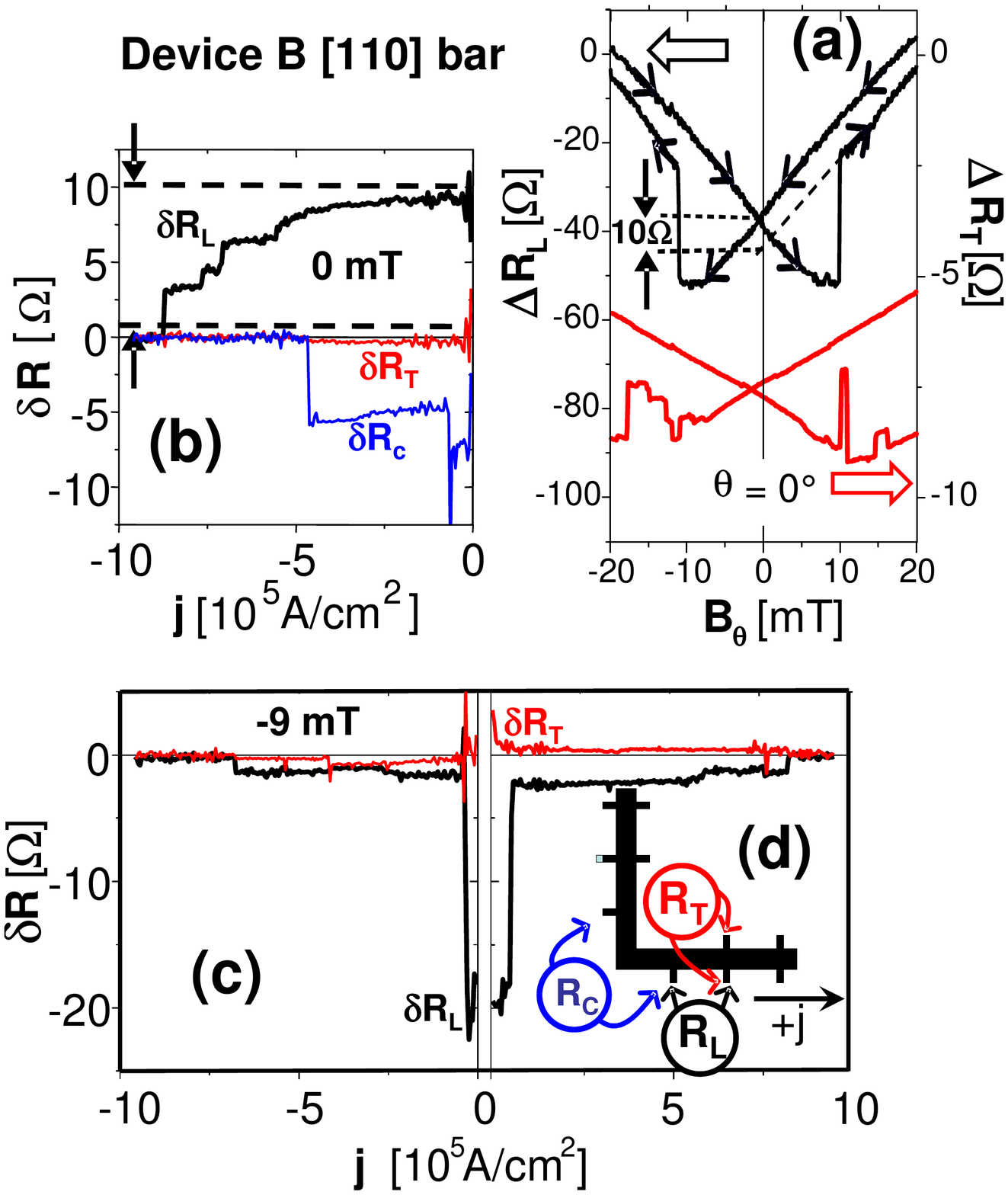}
\vspace{-2cm}
\caption{ (a) Field-sweep measurements at
$\theta=0^{\circ}$  in the [110]-bar of device~B. (b) Difference
between the first and second negative current ramps in the [110]-bar
of device~B at zero field. (c) Difference between the first and
second negative current ramps  at -9~mT field applied along
$\theta=0^{\circ}$. (d) Same as (c)  at for positive current ramps.
The inset shows contacts used for measurements of $R_L$ and $R_T$ in
all panels.} \label{figure9}
\end{figure}

 The distinct
 current induced switching characteristics achieved by patterning one  bar along the [110]
  direction and the other bar along the [1$\overline{1}$0] direction
 are illustrated in Figs.~\ref{figure8} and \ref{figure9} on a set
 of experiments in device~B. The measurements shown in
 Figs.~\ref{figure8}(b)-(d) were taken on the [1$\overline{1}$0]-bar in an external field of a magnitude of -9~mT
 applied along $\theta=0^{\circ}$ (see corresponding field sweep
 measurements in Fig.~\ref{figure8}(a)). Up to the highest experimental current densities, the switching (from
 magnetization angle 9$^{\circ}$ to 180$^{\circ}$) is observed only
 for the positive current polarity. A less detailed tracking of the domain wall is possible in this experiment
 compared to the data in Fig.~\ref{figure7} due to the larger magnitude of the external field
 (larger coercive field of
 device B) and smaller separation of the contacts used to monitor $R_C$ in this
 device. Nevertheless, the -9~mT field assisted reversal process shown in
 Fig.~\ref{figure8} is clearly initiated in the corner and,
 again, the domain wall propagates in the direction opposite to the
 applied hole current. Since for the opposite  magnetic field sweep we observe the
 current induced switching at +9~mT also at positive currents (compare
 Figs.~\ref{figure8}(b) and (d)),
the Oersted fields are unlikely to be the dominant switching
mechanism. Note also that the Oersted fields generated by our
experimental currents are estimated to be two orders of magnitude
weaker than the anisotropy fields.\cite{Yamanouchi:2006_a}

 The character of the current induced switching in device B at -9~mT is completely different in the [110]-bar
 compared to the [1$\overline{1}$0]-bar, as shown in
 Figs.~\ref{figure9}(c) and (d). The switching occurs at much lower current densities due to
 the lower coercive field of the [110]-bar at $\theta=0^{\circ}$
 (compare Figs.~\ref{figure8}(a) and \ref{figure9}(a)), and the
 asymmetry between the positive and negative switching currents is
 small, suggesting that heating plays an important role in this experiment.
 Although we see clear jumps in $\delta R_L$, which are
 consistent with the field-sweep data in Fig.~\ref{figure9}(a), the
 absence of the  $\delta R_T$ switching signal in the [110]-bar
 hinders the unambiguous determination of the switching angles. This
 feature is ascribed to a fabrication induced strong pinning at the
 $R_T$ contacts; indeed the field-sweep measurements for the
 [110]-bar show an incomplete switching at 10~mT in the longitudinal
 resistance and no clear signature of switching for the transverse
 resistance contacts at this field. (Full saturation of the entire
 bar including the transverse contacts region is achieved at
 100~mT.)

 In Fig.~\ref{figure9}(b) we exploit the pinning at the $R_T$
 contacts to study current induced switching at zero magnetic field.
 Note that if the switching of the whole bar was complete the
 zero-field 180$^{\circ}$ rotation from negative to positive
 easy-axis directions would be undetectable by the AMR measurement.
 We again see no switching signal in $\delta R_T$ but a clear step
 in $\delta R_L$. As for all field-assisted experiments, the sense
 and magnitude of the  jump in $\delta R_L$ for zero field
 correlates well with the field sweep measurements (see the dashed
 line in Fig.~\ref{figure9}(a)). Also consistent with the trends in
 the
 field-assisted experiments, the switching occurs at larger current
 than in the -9~mT field assisted switching.  Up to the highest experimental
 current density, the zero-field switching is observed only in the negative current ramp, as we would
 expect for the domain wall propagation from the corner (see the $\delta R_c$ signal in Fig.~\ref{figure9}(b)) to the [110]-bar due to
 spin-transfer-torque. We emphasize however that a detailed
 understanding of the origin of the observed current induced
 switchings in our  L-shaped devices is beyond the scope of this
 work. Our main aim was to demonstrate that the local control of the
 magnetocrystalline anisotropy we achieved in these dilute moment ferromagnetic
 structures is a new powerful
 tool for investigating spin dynamics phenomena.

\section{Summary}
\label{summary} In summary, (Ga,Mn)As microchannels with locally
controlled magnetocrystalline anisotropies and  inherently weak
dipolar fields represent a new favorable class of systems for
exploring magneto-electronic effects at microscale. We have observed
easy-axes rotations which depend on the width and crystal
orientation of the microchannel. Based on numerical simulations of
strain distribution for the experimental geometry and microscopic
calculations of the corresponding spin-orbit coupled band structures
we have explained the effect in terms of lattice relaxation induced
changes in the magnetocrystalline anisotropy. The observation and
explanation of micropatterning controlled magnetocrystalline
anisotropy of the (Ga,Mn)As dilute moment ferromagnet represents the
central result of our paper. In addition to that we have
demonstrated that the structures are well suited for a systematic
study of current induced switching phenomena well bellow Curie
temperature at relatively low current densities. We have found
indications that domain-wall spin-transfer-torque effects contribute
strongly to the observed switchings. This suggests that our
structures represent a new favorable system for exploring these
technologically important yet still physically controversial spin
dynamics phenomena.

{\em Note added}: After the completion of our work, independent and
simultaneous  studies of the lattice relaxation induced changes of
magnetocrystalline anisotropies in (Ga,Mn)As have been posted on the
Los Alamos
Archives and some of them published during the processing of our manuscript.\cite{Humpfner:2006_a,Pappert:2007_a,Wenisch:2007_a} The
crystal orientations and widths of the nanochannels considered in
these works are different than in our study. Nevertheless, the
reported effects are of the same origin and our works provide a
mutual confirmation that the seemingly tiny changes in the lattice
constant can completely overwrite the magnetocrystalline energy
landscape of the host (Ga,Mn)As epilayer.
\section*{Acknowledgment}
We acknowledge discussions with A. H. MacDonald, V. Nov\'ak, and
support from  EU Grant  IST-015728, from EPSRC Grant GR/S81407/01,
from GACR and AVCR  Grants 202/05/0575, 202/06/0025, 202/04/1519,
FON/06/E002, AV0Z1010052, LC510, from MSM Grant 0021620834, from NSF
Grant DMR-0547875, and from ONR Grant N000140610122.


\end{document}